\documentclass[aps,prb,twocolumn,groupedaddress,showpacs,amsmath,amssymb,amsfonts]{revtex4}

\begin{document}
\title{Ground State Fidelity in One Dimensional Gapless Models}

\author{Min-Fong Yang}
\affiliation{Department of Physics, Tunghai University, Taichung,
Taiwan}

\date{\today}

%%%%%%%%%%%%%%%%%%%%%%%%%%%%%%%%%%%%%%%%%%%%%%%%%%%%%%%%%%%%%%%%%%%
\begin{abstract}
A general relation between quantum phase transitions and the
second derivative of the fidelity (or the ``fidelity
susceptibility") is proposed. The validity and the limitation of
the fidelity susceptibility in characterizing quantum phase
transitions is thus established. Moreover, based on the
bosonization method, general formulas of the fidelity and the
fidelity susceptibility are obtained for a class of
one-dimensional gapless systems known as the Tomonaga-Luttinger
liquid. Applying these formulas to the one-dimensional spin-$1/2$
$XXZ$ model, we find that quantum phase transitions, even of the
Beresinskii-Kosterlitz-Thouless type, can be signaled by the
fidelity susceptibility.
\end{abstract}
%%%%%%%%%%%%%%%%%%%%%%%%%%%%%%%%%%%%%%%%%%%%%%%%%%%%%%%%%%%%%%%%%%%


\maketitle

%%%%%%%%%%%%%%%%%%%%%%%%%%%%%%%%%%%%%%%%%%%%%%%%%%%%%%%%%%%%%%%%%%
%\textit{Introduction}.--
%%%%%%%%%%%%%%%%%%%%%%%%%%%%%%%%%%%%%%%%%%%%%%%%%%%%%%%%%%%%%%%%%%

In order to obtain fresh insight into the quantum many-body
problem, a great deal of effort has been devoted to the
application of the concepts emerged from quantum information
science~\cite{Nielsen:book} to the analysis of quantum phase
transitions (QPTs).~\cite{Sachdev:book} QPTs are characterized by
the drastic change in the ground states of quantum many-body
systems, driven solely by quantum fluctuations. Since entanglement
measures the strength of quantum correlations between subsystems
of a compound system, it is natural to expect that entanglement
will be a reliable indicator of QPTs. Much attention has been
focused on the role of entanglement in characterizing QPTs in the
last few years (e.g.
Refs.~\onlinecite{QIS-QPT1,WuSarandyLidar04,deOliveira06,Amico0703044}
and references therein). More recently, another approach to
identify QPTs based on the ground-state fidelity has been
proposed~\cite{Zanardi06} and applied to various many-body
systems.~\cite{ZCG0606130,Cozzini07,CIZ0611727,Buonsante07,Oelkers07,ZGC0701061,CVZ0705.2211,ZCVG0707.2772,YLG0701077,CWGW0706.0072,GKNL0706.2495,zhou,TzengYang07}
Because the fidelity is a measure of similarity between states,
one anticipates that the fidelity should drop abruptly at critical
points, as a consequence of the dramatic changes in the structure
of the ground states, regardless of what type of internal order is
present in quantum many-body states. A perhaps more effective
indicator is given by the singularity in the second derivative of
the fidelity (or the so-called ``fidelity
susceptibility").~\cite{Zanardi06,ZCG0606130,Cozzini07,CIZ0611727,Buonsante07,ZGC0701061,CVZ0705.2211,ZCVG0707.2772,YLG0701077,GKNL0706.2495,note1}
The main advantage of this approach lies in the fact that, since
the fidelity is a purely Hilbert-space geometrical quantity, no
\textit{a priori} knowledge of the structure (order parameter,
correlations driving the QPTs, topology, etc.) of the considered
system is required for its use. The fidelity approach has been
examined in various systems, including Dicke
model,~\cite{Zanardi06} one-dimensional $XY$ model in a transverse
field,~\cite{Zanardi06,zhou} general quadratic fermionic
Hamiltonians,~\cite{ZCG0606130,Cozzini07} and Bose-Hubbard
model.~\cite{Buonsante07,Oelkers07} The success in analyzing QPTs
in these models shows the generality of this procedure. The ground
state fidelity is usually difficult to calculate, due to the lack
of knowledge of the exact ground state wavefunctions. Therefore,
investigations so far are restricted to some particular many-body
models. Conceivably, in order to understand its validity and
limitation, a general connection between the fidelity and QPTs is
highly desired.

In this paper, we discuss, in a general framework, how the
fidelity can be related to a QPT characterized by nonanalyticities
in the derivative of the ground state energy. It is found that,
under certain conditions mentioned below, the fidelity
susceptibility is indeed an effective tool in detecting the
critical points of first-order QPTs (1QPTs) and second-order QPTs
(2QPTs), as illustrated before in several concrete models.
However, it fails to determine the order of the transition, and
may not detect higher-order QPTs. We stress that, although the
fidelity susceptibility can not always detect higher-order QPTs,
it is possible to signal the Beresinskii-Kosterlitz-Thouless (BKT)
transition~\cite{BKT} in one-dimensional many-body systems, which
is a QPT of infinite order. To demonstrate this, analytic formulas
of the fidelity and the fidelity susceptibility are derived for
single-component Luttinger model, which describes a large class of
one-dimensional systems possessing a gapless spectrum. The
fidelity susceptibility is shown to be finite in the thermodynamic
limit even for these gapless systems. Besides, it is a smooth
function of the controlling parameter driving QPTs [see
Eq.~(\ref{eq:fide_2_deri}) below], as long as the system lying in
the critical Luttinger-liquid phase. By using these formulas, the
fidelity and the fidelity susceptibility can be easily calculated
for a large class of gapless systems, as long as the relation
between the Luttinger liquid parameters and the controlling
parameter driving QPTs is known. Employing our general formulas to
the one-dimensional spin-$1/2$ $XXZ$ model, we show that the BKT
transition therein can indeed be signaled by the singularity in
the fidelity susceptibility. Though we restrict our attention to
the $XXZ$ spin chain, according to the general expressions of the
fidelity and the fidelity susceptibility, one can expect that same
results should apply to all the BKT-type QPTs of one-dimensional
models.

%%%%%%%%%%%%%%%%%%%%%%%%%%%%%%%%%%%%%%%%%%%%%%%%%%%%%%%%%%%%%%%%%%
%\textit{QPT and the fidelity}.--
%%%%%%%%%%%%%%%%%%%%%%%%%%%%%%%%%%%%%%%%%%%%%%%%%%%%%%%%%%%%%%%%%%

We now begin to show the general relation between quantum phase
transitions and the fidelity susceptibility. Only QPTs
characterized by nonanalytic behavior in the derivatives of the
ground state energy are considered here. According to the
conventional classification, a 1QPT is characterized by a finite
discontinuity in the first derivative of the ground state energy.
Similarly, a 2QPT is characterized by a finite discontinuity, or
divergence, in the second derivative of the ground state energy,
assuming the first derivative is continuous. Following the
notations in Ref.~\onlinecite{WuSarandyLidar04}, the most general
Hamiltonian of $N$ distinguishable particles governed by up to
two-body interactions can be written as
\begin{equation*}
H=\sum_{i\alpha \beta }\epsilon _{\alpha \beta }^{i}\left\vert
\alpha _{i}\right\rangle \left\langle \beta _{i}\right\vert
+\sum_{ij\alpha \beta \gamma \delta }V_{\alpha \beta \gamma \delta
}^{ij}\left\vert \alpha _{i}\right\rangle \left\vert \beta
_{j}\right\rangle \left\langle \gamma _{i}\right\vert \left\langle
\delta _{j}\right\vert \; ,
\end{equation*}
where $\{\left\vert \alpha_{i}\right\rangle \}$ is a basis for the
local Hilbert space of particle $i$. For these systems, it has
been shown that the derivatives of the ground state energy per
particle $\mathcal{E}$ can be written as~\cite{WuSarandyLidar04}
\begin{eqnarray}
\frac{\partial\mathcal{E}}{\partial\lambda} &=&
\frac{1}{N}\sum_{ij}
\mathrm{tr}\left(\frac{\partial\mathbf{U}(ij)}{\partial\lambda}\rho^{ij}\right)
\; , \label{der 1 E} \\
\frac{\partial^{2}\mathcal{E}}{\partial\lambda^{2}}
&=&\frac{1}{N}\sum_{ij}\left[ \mathrm{tr}
\left(\frac{\partial^{2}\mathbf{U}(ij)}{\partial\lambda^{2}}\rho^{ij}\right)
+ \mathrm{tr} \left(
\frac{\partial\mathbf{U}(ij)}{\partial\lambda} \frac{\partial\rho
^{ij}}{\partial\lambda} \right) \right] \; , \label{der 2 E}
\end{eqnarray}
where $\lambda$ is a controlling parameter of the system's
Hamiltonian and $\mathrm{tr}$ denotes the trace over the degrees
of freedom for two particles. $\mathbf{U}(ij)$ denotes a matrix
with matrix elements $U_{\alpha \beta ,\gamma \delta
}(ij)=\epsilon _{\alpha \gamma }^{i}\delta _{\beta \delta
}^{j}/N_i+V_{\alpha \beta \gamma \delta }^{ij}$, where $\delta
_{\beta \delta }^{j}$ is the Kronecker symbol on particle $j$, and
$N_i$ is the number of particles that particle $i$ interacts with.
$\rho^{ij}= \mathrm{Tr}_{\hat{i}\hat{j}}\;\rho_0(\lambda)$ is the
two-particle reduced density matrix, which is obtained by taking a
partial trace $\mathrm{Tr}_{\hat{i}\hat{j}}$ over all degrees of
freedom except particles $i$ and $j$. Here $\rho_0(\lambda)\equiv
|\Psi_0 (\lambda)\rangle \langle\Psi_0 (\lambda)|$ is the density
matrix of the ground state with $|\Psi_0 (\lambda)\rangle$ being
the normalized ground state. $\mathbf{U}(ij)$ includes all the
single and two-body terms of the Hamiltonian associated with
particles $i$ and $j$. We assume that $\mathbf{U}(ij)$ is a smooth
function of the Hamiltonian parameter $\lambda$. From
Eqs.~(\ref{der 1 E}) and (\ref{der 2 E}), one finds that the
origin of a 1QPT (2QPT) is due to the fact that one or more of the
matrix elements of $\rho^{ij}$ ($\partial_{\lambda}\rho^{ij}$) are
discontinuous/divergent at the transition point
$\lambda=\lambda_c$.~\cite{WuSarandyLidar04}

The quantum fidelity (or the modulus of the overlap) $F$ of two
normalized ground states $|\Psi_0 (\lambda+\delta\lambda)\rangle$
and $|\Psi_0 (\lambda)\rangle$ corresponding to neighboring
Hamiltonian parameters is given by
$F(\lambda+\delta\lambda,\lambda) = |\langle\Psi_0
(\lambda+\delta\lambda)|\Psi_0
(\lambda)\rangle|$.~\cite{Zanardi06} To detect QPTs, a more
effective indicator is provided by the peak in the ``density'' of
the second derivative of the fidelity  (or the so-called
``fidelity susceptibility") ${\cal
S}(\lambda)$,~\cite{Zanardi06,ZCG0606130,Cozzini07,CIZ0611727,Buonsante07,ZGC0701061,CVZ0705.2211,ZCVG0707.2772,YLG0701077,GKNL0706.2495,note1}
which is free from the arbitrariness of small parameter
$\delta\lambda$. In the thermodynamic limit (i.e., both the
particle number $N$ and the number of lattice sites $L$ approach
to infinity, while $N/L$ keeps finite), ${\cal S}(\lambda)$ can be
written as~\cite{note2}
\begin{equation} \label{def:S}
{\cal S}(\lambda)= \lim_{\delta\lambda \to 0} \lim_{L \rightarrow
\infty} \frac{-2 \ln
F(\lambda+\delta\lambda,\lambda)}{L\;\delta\lambda^2} \; .
\end{equation}
To make comparison with the expressions of the derivatives of the
ground state energy, we rewrite the quantum fidelity and the
fidelity susceptibility in terms of the density matrix
$\rho_0(\lambda)$ of the ground state. Notice that
\begin{equation}
F(\lambda+\delta\lambda,\lambda)^2 = \mathrm{Tr}
[\rho_0(\lambda)\rho_0(\lambda+\delta\lambda)] \; .
\end{equation}
Now expanding $\rho_0(\lambda+\delta\lambda)$ in powers of
$\delta\lambda$, and using the identity $\partial_\lambda
[\langle\Psi_0(\lambda)|\Psi_0(\lambda)\rangle]=0$ which implies
$\mathrm{Tr}[\rho_0(\lambda)\partial_\lambda\rho_0(\lambda)]=0$,
one can easily show that
\begin{equation}
F(\lambda+\delta\lambda,\lambda) \simeq 1 -
\frac{\delta\lambda^2}{4} \mathrm{Tr} \left[
\frac{\partial\rho_0(\lambda)}{\partial\lambda}
\frac{\partial\rho_0(\lambda)}{\partial\lambda} \right] \; ,
\label{F}
\end{equation}
and thus
\begin{equation}
{\cal S}(\lambda)= \lim_{L \rightarrow \infty} \frac{1}{2L}
\mathrm{Tr} \left[ \frac{\partial\rho_0(\lambda)}{\partial\lambda}
\frac{\partial\rho_0(\lambda)}{\partial\lambda} \right] \; .
\label{delF}
\end{equation}
As mentioned before, 1QPTs (2QPTs) must come from discontinuity in
(discontinuity in or divergence of) one or more matrix elements of
$\rho^{ij}$ ($\partial_{\lambda}\rho^{ij}$). Since the matrix
elements of the reduced density matrix $\rho^{ij}$ are linear
functions of those of $\rho_0$, 1QPTs and 2QPTs will be associated
with nonanalyticity in the matrix elements of
$\partial_{\lambda}\rho_0$, and therefore imply the singularity in
${\cal S}(\lambda)$. That is, the singular behavior of the
fidelity susceptibility is able to signal 1QPTs and 2QPTs. This
explains the success of the fidelity approach discovered in the
previous investigations. \textit{Note that the above conclusion is
valid only if the discontinuous/divergent quantities do not
accidentally all vanish or cancel with other terms in
Eqs.~(\ref{F}) and (\ref{delF})} [i.e., assumptions (b) and (c) in
Ref.~\onlinecite{WuSarandyLidar04}]. However, some limitations of
this approach are discussed in order. First, even though ${\cal
S}(\lambda)$ can be a good indicator of 1QPTs and 2QPTs, it fails
to distinguish between them, in contrast to the entanglement
measurements discussed in
Refs.~\onlinecite{WuSarandyLidar04,deOliveira06}. Second, from
Eq.~(\ref{delF}), we find that ${\cal S}(\lambda)$ can not detect
the higher-order QPTs resulting from the nonanalyticities in
$\partial_{\lambda}^{2}\rho_0$ and even higher-order derivatives.
Nevertheless, it does not mean that ${\cal S}(\lambda)$ always
fails to signal the higher-order QPTs. Reminding that the
two-particle reduced density matrix $\rho^{ij}$ is calculated by
taking a partial trace of $\rho_0$, it is thus possible that the
nonanalyticities in the matrix elements of
$\partial_{\lambda}\rho_0$ cancel one another in obtaining
$\partial_{\lambda}\rho^{ij}$. That is, while the
discontinuity/divergence in $\rho^{ij}$ and
$\partial_{\lambda}\rho^{ij}$ does imply the nonanalyticities in
$\partial_{\lambda}\rho_0$, the reverse is not true. In this case,
$\partial_{\lambda}^{2}\mathcal{E}$ can be continuous even though
${\cal S}(\lambda)$ is singular, and therefore such a higher-order
QPT can indeed be detected by ${\cal S}(\lambda)$. One such
example is provided by the BKT transition in the one-dimensional
spin-$1/2$ $XXZ$ model, where a critical anisotropy separates a
gapless phase from a gapful phase. As demonstrated in
Ref.~\onlinecite{Yang}, the ground state energy and all of its
derivatives with respect to the anisotropy are continuous at the
critical point. That is, it is a QPT of infinite order. However,
as discussed below, ${\cal S}(\lambda)$ does become singular
despite the regularity of the ground state energy at this critical
point.

%%%%%%%%%%%%%%%%%%%%%%%%%%%%%%%%%%%%%%%%%%%%%%%%%%%%%%%%%%%%%%%%%%
%\textit{one-dimensional systems in their Luttinger liquid phase}.--
%%%%%%%%%%%%%%%%%%%%%%%%%%%%%%%%%%%%%%%%%%%%%%%%%%%%%%%%%%%%%%%%%%

It is known that many one-dimensional gapless systems, which may
undergo the BKT transition, can be described by a single-component
Luttinger model.~\cite{Giamarchi:book} They include the spin-$1/2$
$XXZ$ spin chain and the spin-$1/2$ $J-J^\prime$ spin chain in
their spin fluid phases, and the Bose-Hubbard model in its
superfluid phase, etc. Before specifying to the $XXZ$ spin chain,
we first derive analytic formulas of the fidelity and the fidelity
susceptibility for the single-component Luttinger model. By using
the bosonization method, the Luttinger model can be described by
the following effective Hamiltonian~\cite{Giamarchi:book,note3}
\begin{equation}\label{hbos0}
H_\textrm{eff} = \frac{u}{2}  \int dx \left[ K \Pi(x)^2 +
\frac{1}{K} (\partial_x \Phi )^2 \right] \; .
\end{equation}
Here the bosonic phase field operators $\Phi$ and $\Pi$ are given
by~\cite{note4}
\begin{eqnarray}
\Phi(x) & \sim & - \frac{i\sqrt{\pi}}{L} \sum_{k \neq 0}
\left(\frac{L|k|}{2\pi}\right)^{1/2} \frac{1}{k}\; e^{-ikx}
\left(a^\dag_k + a_{-k}\right)
\; , \nonumber \\
\Pi(x) & \sim & \frac{\sqrt{\pi}}{L}\sum_{k \neq 0}
\left(\frac{L|k|}{2\pi}\right)^{1/2} \frac{k}{|k|}\; e^{-ikx}
\left(a^\dag_k - a_{-k}\right) \; , \nonumber
\end{eqnarray}
where $a_k$ and $a^\dag_{k^\prime}$ obey canonical boson
commutation relations: $[a_k, a^\dag_{k^\prime}] = \delta_{k,
k^\prime}$. We note that the Luttinger liquid parameters $K$ and
$u$ are functions of the controlling parameter $\lambda$ of the
original Hamiltonian. When $K=1$, the effective Hamiltonian
reduces to the free-boson model, whose ground state
$|\Psi_0(K=1)\rangle$ is nothing but the vacuum $|0\rangle$ of the
canonical bosons satisfying $a_{k} |0\rangle =0$ for all $k \neq
0$. When $K \neq 1$, the effective Hamiltonian contains the
unpleasant terms $a_k a_{-k}$ and $a^\dag_k a^\dag_{-k}$. These
terms can be diagonalized by a Bogoliubov
transformation on the bosons such that
\begin{eqnarray}
\alpha_{-k} &=& \cosh\theta\; a_{-k} + \sinh\theta\; a^\dag_k   \nonumber \\
\alpha^\dag_k &=& \sinh\theta\; a_{-k} + \cosh\theta\; a^\dag_k
\nonumber
\end{eqnarray}
with $\cosh\theta=(1+K)/2\sqrt{K}$ and
$\sinh\theta=(1-K)/2\sqrt{K}$. We note that, in the present case,
the parameter $\theta$ is independent of the momentum $k$. The
transformed Hamiltonian becomes a free-boson model in terms of the
new set of canonical bosons $\alpha_k$. That is, the effective
model in Eq.~(\ref{hbos0}) can be considered as a ``quasi-free"
boson model, where the ground states $|\Psi_0(K)\rangle$ for
general values of $K$ are the vacuum of $\alpha_k$ satisfying
$\alpha_{k} |\Psi_0(K)\rangle =0$ for all $k \neq 0$. Thus the
(not normalized) ground states become
\begin{equation} \label{eq:GS}
|\Psi_0(K)\rangle = \exp \left( -\frac{\sinh\theta}{\cosh\theta}
\sum_{k \neq 0} a^\dag_k a^\dag_{-k} \right) |0\rangle \; .
\end{equation}

With these exact expressions of the ground states, now one can
calculate the ground-state fidelity. From its definition, the
ground-state fidelity $F$ of two (not normalized) ground states
$|\Psi_0 (K^\prime)\rangle$ and $|\Psi_0 (K)\rangle$ becomes
$F(K^\prime,K) = Z(K^\prime,K)/\sqrt{Z(K^\prime,K^\prime)Z(K,K)}$,
where $Z(K^\prime,K) \equiv |\langle\Psi_0 (K^\prime)|\Psi_0
(K)\rangle|$. By using the expressions of the ground states in
Eq.~(\ref{eq:GS}), it can be shown that
\begin{equation}
Z(K^\prime,K) = \prod_{k \neq 0} \left(1-\frac{\sinh\theta^\prime
\sinh\theta}{\cosh\theta^\prime \cosh\theta} \right)^{-1} \; ,
\end{equation}
and therefore a general expression of the fidelity is reached,
\begin{equation}
F(K^\prime,K) = \prod_{k \neq 0} \frac{1}{\cosh(\theta^\prime
-\theta)} = \prod_{k \neq 0} \frac{2}{\sqrt{\frac{K}{K^\prime}} +
\sqrt{\frac{K^\prime}{K}}} \; . \label{eq:F}
\end{equation}
Here the prime denotes the corresponding variables taking their
values at $K^\prime$. Obviously, $F=1$ if $K^\prime=K$.
Generically, $1/\cosh(\theta^\prime-\theta) < 1$. Therefore, for
systems with a large but finite size $L$, the fidelity in
Eq.~(\ref{eq:F}) scales as $[\cosh(\theta^\prime-\theta)]^{-L}$,
and decays very fast when $K^\prime$ separates from $K$. That is,
the fidelity of two different ground states becomes zero in the
thermodynamic limit, despite the two states being in the same
phase. Therefore, in this case, it is difficult to signal a
precursor of the QPT simply by seeking a drop in the fidelity. As
mentioned before, a more effective indicator is provided by the
fidelity susceptibility given in Eq.~(\ref{def:S}). From the
analytic expression of the fidelity in Eq.~(\ref{eq:F}), it can be
shown that
\begin{equation}
\lim_{L \rightarrow \infty} \frac{\ln F(K+\delta K,K)}{L} \simeq
-\frac{1}{8} \left( \frac{1}{K}\frac{dK}{d\lambda} \right)^2
\delta\lambda^2 \; ,
\end{equation}
and one finally obtains the general formula of the fidelity
susceptibility
\begin{equation}
{\cal S}(\lambda) = \frac{1}{4} \left(
\frac{1}{K}\frac{dK}{d\lambda} \right)^2 \; .
\label{eq:fide_2_deri}
\end{equation}
Thus we show that ${\cal S}(\lambda)$ can be finite in the
thermodynamic limit even for gapless systems. This result agrees
with the findings in Ref.~\onlinecite{CVZ0705.2211} based on the
scaling arguments. In the present work, we further provide the
analytic expression of ${\cal S}(\lambda)$. From this expression,
we find that ${\cal S}(\lambda)$ becomes singular only when
$dK/d\lambda$ diverges. It is noted that, in terms of the
Luttinger liquid parameter $K$, the general expressions of $F$ and
${\cal S}(\lambda)$ in Eqs.~(\ref{eq:F}) and
(\ref{eq:fide_2_deri}) apply to all one-dimensional systems in
their Luttinger liquid phases. The precise location of the
singularity in ${\cal S}(\lambda)$ can be determined, once the
exact relation between the Luttinger liquid parameter $K$ and the
controlling parameter $\lambda$ driving QPTs is known.

%%%%%%%%%%%%%%%%%%%%%%%%%%%%%%%%%%%%%%%%%%%%%%%%%%%%%%%%%%%%%%%%%%
%\textit{$XXZ$ spin chain}.--
%%%%%%%%%%%%%%%%%%%%%%%%%%%%%%%%%%%%%%%%%%%%%%%%%%%%%%%%%%%%%%%%%%

Now we consider the case of the $XXZ$ spin chain, which can be
taken as a special case in the Luttinger liquid description. The
Hamiltonian of the one-dimensional spin-$1/2$ $XXZ$ model is
written by
\begin{equation} \label{eq:XXZ}
H = \sum_{j=1}^L (S ^x_j  S ^x_{j+1} +  S ^y_j  S ^y_{j+1} +
\lambda S ^z_j  S ^z_{j+1}) \; .
\end{equation}
Here $S_j^x, S_j^y$, and $S_j^z$ are the spin-$1/2$ operators at
the $j$-th lattice site. The parameter $\lambda$ denotes an
anisotropy in the spin-spin interaction. The $XXZ$ spin chain is
an exactly solvable model.~\cite{Yang} It is known that there is a
critical point of 1QPT at $\lambda=-1$, which corresponds to the
isotropic ferromagnetic Heisenberg model. At the isotropic
antiferromagnetic point $\lambda=1$, a BKT transition occurs,
which is described by a divergent correlation length but without
true long-range order.

After applying the Jordan-Wigner transformation and bosonization
procedure for the spin-1/2 operators, when $-1 < \lambda \leq 1$,
the Hamiltonian in Eq.~(\ref{eq:XXZ}) can be mapped to the
bosonized effective Hamiltonian in
Eq.~(\ref{hbos0}).~\cite{Giamarchi:book} That is, the $XXZ$ spin
chain can be considered as a ``quasi-free" boson model. For
general values of $\lambda$ (obeying $-1 < \lambda \leq 1$), the
Luttinger liquid parameters $K$ and $u$ can be obtained by
comparison with the Bethe-Ansatz solution. They are given by
$K=(\pi/2)/[\pi-\arccos(\lambda)]$ and
$u=\pi\sqrt{1-\lambda^2}/(2\arccos\lambda)$.~\cite{Luttinger_parameters}
For $\lambda = 0$, we have $K = 1$; while $K \neq 1$ for $\lambda
\neq 0$. Substituting the above exact relation between the
Luttinger liquid parameter $K$ and the anisotropy parameter
$\lambda$ to Eqs.~(\ref{eq:F}) and (\ref{eq:fide_2_deri}), the
expressions of $F(\lambda^\prime,\lambda)$ and ${\cal S}(\lambda)$
can be obtained. Here we discuss ${\cal S}(\lambda)$ only, which
becomes
\begin{equation}
{\cal S}(\lambda) = \frac{1}{4 [\pi-\arccos(\lambda)]^2}
\frac{1}{1-\lambda^2}\; . \label{eq:fide_2_deri_XXZ}
\end{equation}
Therefore, the fidelity susceptibility ${\cal S}(\lambda)$
diverges as $\lambda \rightarrow \pm 1$. That is, the singular
behavior in ${\cal S}(\lambda)$ is able to signal either the 1QPT
at $\lambda=-1$ or the BKT transition at $\lambda=1$. According to
our previous analysis, the singularity in ${\cal S}(\lambda)$ at
$\lambda=1$ indicates that one or more of the matrix elements of
$\partial_{\lambda}\rho_0$ should be divergent at this BKT
transition. Since the BKT transition is a QPT of infinite order,
nonanalyticities in the density matrix of ground state $\rho_0$
must accidentally all vanish or cancel with other terms, such that
the ground state energy and all of its derivatives with respect to
the anisotropy are continuous at this critical point. Therefore,
the BKT transition in the $XXZ$ spin chain does provide an
example, where the higher-order QPTs can be detected by the
singularity in the fidelity susceptibility.

%%%%%%%%%%%%%%%%%%%%%%%%%%%%%%%%%%%%%%%%%%%%%%%%%%%%%%%%%%%%%%%%%%
%\textit{conclusion}.--
%%%%%%%%%%%%%%%%%%%%%%%%%%%%%%%%%%%%%%%%%%%%%%%%%%%%%%%%%%%%%%%%%%
In summary, according to the proposed general relation between
QPTs and the fidelity susceptibility, the validity and the
limitation of the fidelity susceptibility in characterizing QPTs
are discussed. Employing our analytic formulas of the fidelity and
the fidelity susceptibility for the one-dimensional Luttinger
model, it is shown that the fidelity susceptibility can be finite
even for critical systems, which agrees with the result in
Ref.~\onlinecite{CVZ0705.2211} based on the scaling analysis.
Moreover, while the fidelity susceptibility may not detect
higher-order QPTs in general, we demonstrate that the BKT
transition, a QPT of infinite order, in the spin-1/2 $XXZ$ chain
can indeed be signaled by the singularity in the fidelity
susceptibility. Though we restrict our attention to the $XXZ$ spin
chain, we believe that same results will apply to all the BKT-type
QPTs of one-dimensional models, such as the transition from spin
fluid to dimerized phase in the $J-J^\prime$ model, and the
superfluid-insulator transition in the Bose-Hubbard model at
integer filling, etc.

%%%%%%%%%%%%%%%%%%%%%%%%%%%%%%%%%%%%%%%%%%%%%%%%%%%%%%%%%%%%%%
%\begin{acknowledgments}
The author is grateful to M.-C. Chang for many valuable
discussions. This work was supported by the National Science
Council of Taiwan under Contract No. NSC 96-2112-M-029-004-MY3.
%\end{acknowledgments}
%%%%%%%%%%%%%%%%%%%%%%%%%%%%%%%%%%%%%%%%%%%%%%%%%%%%%%%%%%%%%%


\begin{thebibliography}{21}


%%%%%% A book on QIS %%%%%%
\bibitem{Nielsen:book}
M. A. Nielsen and I. L. Chuang, {\it Quantum Computation and
Quantum Information} (Cambridge University Press, Cambridge,
2000).


%%%%%% A book on QPTs %%%%%%
\bibitem{Sachdev:book}
S. Sachdev, {\it Quantum Phase Transitions}, (Cambridge University
Press, Cambridge, 1999).


%%%%%%  transverse field Ising chains  %%%%%%
\bibitem{QIS-QPT1}
A. Osterloh, L. Amico, G. Falci, and R. Fazio,
%A. Osterloh {\it et al.},
Nature \textbf{416}, 608 (2002); T. J. Osborne and M. A. Nielsen,
Phys. Rev. A \textbf{66}, 032110 (2002); Quantum Inf. Process.
{\bf 1}, 45 (2002).


\bibitem{WuSarandyLidar04}
L.-A. Wu, M. S. Sarandy, and D. A. Lidar,
%L.-A. Wu  {\it et al.},
Phys. Rev. Lett. \textbf{93}, 250404 (2004).



\bibitem{deOliveira06}
T. R. de Oliveira, G. Rigolin, M. C. de Oliveira, and E. Miranda,
%T. R. de Oliveira {\it et al.},
Phys. Rev. Lett. \textbf{97}, 170401 (2006).


%%%%%%  review   %%%%%%
\bibitem{Amico0703044}
For a recent review, see L. Amico, R. Fazio, A. Osterloh, and J.V.
Vedral,
%L. Amico {\it et al.},
quant-ph/0703044.


%%%%%% fidelity: Dicke and XY models %%%%%%
\bibitem{Zanardi06}
P. Zanardi and N. Paunkovi\'{c}, Phys. Rev. E \textbf{74}, 031123
(2006).


%%%%%% fidelity & QPT in free fermion systems %%%%%%
\bibitem{ZCG0606130}
P.~Zanardi, M.~Cozzini, and P.~Giorda,
%P. Zanardi {\it et al.},
J. Stat. Mech. (2007) L02002.
%arXiv: quant-ph/0606130.


%%%%%% fidelity & QPT in free fermion graphs %%%%%%
\bibitem{Cozzini07}
M.~Cozzini, P.~Giorda, and P.~Zanardi,
%M.~Cozzini {\it et al.},
Phys. Rev. B \textbf{75}, 014439 (2007).
%(arXiv: quant-ph/0608059).


%%%%%% fidelity & QPT in matrix-product states %%%%%%
\bibitem{CIZ0611727}
M.~Cozzini, R.~Ionicioiu, and P.~Zanardi,
%M. Cozzini {\it et al.},
Phys. Rev. B \textbf{76}, 104420 (2007).
%arXiv: cond-mat/0611727


%%%%%% fidelity & QPT in Bose-Hubbard Model %%%%%%
\bibitem{Buonsante07}
P. Buonsante and A. Vezzani, Phys. Rev. Lett. \textbf{98}, 110601
(2007).


%%%%%% fidelity in Bose-Hubbard Model using Bethe ansatz %%%%%%
\bibitem{Oelkers07}
N. Oelkers and J. Links, Phys. Rev. B \textbf{75}, 115119 (2007).


%%%%%% Riemannian metric tensor & QPT %%%%%%
\bibitem{ZGC0701061}
P. Zanardi, P. Giorda, and M. Cozzini,
%P. Zanardi {\it et al.},
Phys. Rev. Lett. \textbf{99}, 100603 (2007).
%arXiv: quant-ph/0701061v1
\bibitem{CVZ0705.2211}
L. Campos Venuti and P. Zanardi, Phys. Rev. Lett. \textbf{99},
095701 (2007).
%arXiv:0705.2211.
\bibitem{ZCVG0707.2772}
P. Zanardi, L. Campos Venuti, and P. Giorda,
%P. Zanardi {\it et al.},
arXiv:0707.2772.


%%%%%% fidelity susceptibility-1 %%%%%%
\bibitem{YLG0701077}
W. L. You, Y. W. Li, and S. J. Gu,
%W. L. You {\it et al.},
Phys. Rev. E \textbf{76}, 022101 (2007).
%arXiv: quant-ph/0701077.


%%%%%% fidelity susceptibility-2 %%%%%%
\bibitem{GKNL0706.2495}
S. J. Gu, H. M. Kwok, W. Q. Ning, and H. Q. Lin,
%S. J. Gu {\it et al.},
arXiv:0706.2495.


%%%%%% fidelity & QPT in S=1/2 J-J' Model %%%%%%
\bibitem{CWGW0706.0072}
S. Chen, L. Wang, S. J. Gu, and Y. Wang,
%S. Chen {\it et al.},
arXiv:0706.0072.


\bibitem{zhou}
H. Q. Zhou and J. P. Barjaktarevic, arXiv: cond-mat/0701608; H. Q.
Zhou, J. H. Zhao, B. Li,
%H. Q. Zhou {\it et al.},
arXiv:0704.2940; H. Q. Zhou, arXiv:0704.2945.


\bibitem{TzengYang07}
Y. C. Tzeng and  M. F. Yang, arXiv:0709.1518.



\bibitem{note1}
Similar quantities, say the Riemannian metric tensor inherited
from the parameter
space,~\cite{ZGC0701061,CVZ0705.2211,ZCVG0707.2772} have also been
prosposed.


\bibitem{BKT}
V. L. Beresinskii, Sov. Phys. JETP {\bf 32}, 493 (1971); J. M.
Kosterlitz and D. J. Thouless, J. Phys. C \textbf{6}, 1181 (1973);
J. M. Kosterlitz, {\it ibid}. \textbf{7}, 1046 (1974).


\bibitem{note2}
Because $F(\lambda+\delta\lambda, \lambda)$ reaches its maximum
for $\delta\lambda=0$, expanding the fidelity in powers of
$\delta\lambda$, the first derivative vanishes and
$F(\lambda+\delta\lambda, \lambda) \simeq 1 +
[\partial_{\lambda^\prime}^2 F(\lambda^\prime,
\lambda)]_{\lambda^\prime=\lambda} \; \delta\lambda^2/2$.
Therefore, one can show that the expression in Eq.~(\ref{def:S})
is identical to that given in the literature.


%%%%%%  Bethe ansatz foe XXZ model   %%%%%%
\bibitem{Yang}
C. N. Yang  and C. P. Yang, Phys. Rev. {\bf 150}, 321 (1966); {\it
ibid.} {\bf 150}, 327 (1966). %; {\it ibid.} {\bf 151}, 258 (1966).


\bibitem{Giamarchi:book}
T. Giamarchi, {\it Quantum Physics in One Dimension} (Oxford
University Press, New York, 2004).


\bibitem{note3}
Here we use the notations where $\sqrt{\pi}\Pi$ and
$\Phi/\sqrt{\pi}$ employed in Ref.~\onlinecite{Giamarchi:book} are
identified as $\Pi$ and $\Phi$.


\bibitem{note4}
See, for example, Appendix D in Ref.~\onlinecite{Giamarchi:book}.
Here the zero-mode operators are omitted and a cutoff parameter is
dropped for simplicity.



\bibitem{Luttinger_parameters}
J. D. Johnson, S. Krinsky, and B. McCoy,
%J. D. Johnson {\it et al.},
Phys. Rev. A \textbf{8}, 2526 (1973); A. Luther and I. Peschel,
Phys. Rev. B \textbf{12}, 3908 (1975); F. D. M. Haldane, Phys.
Rev. Lett. \textbf{45}, 1358 (1980).


\end{thebibliography}
\end{document}